# Fingering instabilities in binary granular systems

Meng Liu [a], Nicholas A. Conzelmann [a], Louis Girardin [b], Fabian J. Dickhardt [c], Christopher P. McLaren [a], Jens. P. Metzger [a], Christoph R. Müller [a*]

[a] Department of Mechanical and Process Engineering, ETH Zurich, Leonhardstrasse 21,8092 Zurich, Switzerland
[b] Department of Mechanical Engineering, University College of London, W1W 7TS London, UK
[c] Department of Mechanical Engineering, Massachusetts Institute of Technology, USA
* Corresponding author email address: muelchri@ethz.ch

**Abstract**

Fingering instabilities akin to the Rayleigh-Taylor (RT) instability in fluids have been observed in a binary granular system consisting of dense and small particles layered on top of lighter and larger particles, when the system is subjected to vertical vibration and fluidizing gas flow. Using observations from experiments and numerical modelling we explore whether the theory developed to describe the Rayleigh-Taylor (RT) instability in fluids is also applicable to binary granular systems. Our results confirm the applicability of the classic RT instability theory for binary granular systems demonstrating that several key features are observed in both types of systems, viz: (i) The characteristic wavenumber of the instability is constant with time, (ii) the amplitude of the characteristic wavenumber initially grows exponentially and (iii) the dispersion relation between the wavenumbers $k$ of the interface instability and the growth rates $n(k)$ of their amplitudes holds in both fluid-fluid and binary granular systems. Our results also demonstrate that inter-particle friction is essential for the RT instability to occur in granular media. For zero particle friction the interface instability bears a greater resembles to the Richtmyer-Meshkov instability. We further define a yield criterion $Y$ for the interface by treating the granular medium as a viscoplastic material; only for $Y > 15$ fingering occurs. Interestingly, previous work has shown that instabilities in the Earth's lower mantle, another viscoplastic material, also occur for similar values of $Y$.

## 1. Introduction

Rayleigh-Taylor (RT) instabilities arise at the interface between two fluids of different densities; for example, when a fluid with a higher density is layered on top of a fluid of a lower density in a gravitational field. Any perturbation of the (flat) interface induces a local pressure gradient (acting in the normal direction to the interface) leading to a growth of the perturbation, which manifests itself as fingers intruding upwards and downwards into the fluid layers [1]. While the RT instability is a classical phenomenon in fluid mechanics, more recently it has been reported also in granular systems such as granular suspensions [2,3], gas-particle mixtures [4,5], or particle-particle mixtures [6]. Beyond these laboratory systems, RT instabilities are also observed in geological settings, such as plumes rising upwards in the Earth's mantle [7], river bed erosion [8] or the formation of salt domes [9]

For fluid-particle systems in which the forces acting on the particles are dominated by fluid-particle interactions (e.g. drag) while particle inertia is negligible, a hydrodynamic description of the formation of RT-like instabilities has been proposed [10,11] and experimentally verified [3,12]. Typically, in these models the particle-fluid mixture is treated as an immiscible fluid phase with a uniform solid fraction (placed on top of a layer of a pure fluid). Linear stability analysis has shown that the initial growth of the interface in fluid-particle systems can indeed be described by the dispersion relation which was developed originally for the RT instability in fluid-fluid systems [13]. The dispersion relation provides a relationship between the interface perturbations, characterized by a series of wavenumbers $k$, and their respective growth rates $n(k)$. According to Taylor's instability theory, perturbations of the interface grow exponentially with time [1]. A good quantitative agreement between such a linear stability analysis and experimental measurements has been observed for a glycerin-particle system in a Hele-Shaw-like cell where a glycerin-particle suspension was at the bottom while a glycerin layer was placed on top [3]. By inverting the Hele-Shaw cell, the suspension sinks under gravity while glycerin rises up, exhibiting classic fingering structures.

Finger-instabilities have also been observed in a radial Hele-Shaw-like cell in which a fluid penetrates into a single-phase granular medium [14-16]. However, in such a configuration the fluid fingers side-branch rather than split at the tip as it is commonly observed for fingering in Newtonian fluids [17]. This side-branching of the viscous fingers is considered as a granular equivalent of the classic Saffman-Taylor instability in the zero-surface-tension limit. The Saffman-Taylor instability occurs when injecting a viscous fluid into a more viscous one. A quantitative analysis of experiments has shown however that the scaling of the growth of fingers in a granular medium is distinct from the scaling in fluid-fluid systems; in granular systems the finger width $W_{finger}$ follows the scaling $W_{finger} \sim U_i^{1/2}$ [15], while in conventional Newtonian fluids $W_{finger} \sim U_i^{-1/2}$ is observed, where $U_i$ is the local interfacial growth velocity, i.e. the velocity with which fingers grow radially (identified by tracing the outermost boundary of the fingers). This very distinct scaling behaviour is believed to be a consequence of the different dissipation mechanism in granular materials, viz. friction-induced dissipation, as opposed to viscous damping in fluids [18-20].

In granular media interface instabilities have also been observed when a layer of a dense granular material (instead of a liquid) is placed above a layer of air [5]. Here, fingering patterns emerge as the granular material falls under gravity.

This has been reported, e.g. by Vinningland *et al.* [5] who tracked polystyrene particles falling in a Hele-Shaw cell. Vinningland *et al.* [5] determined the interface wavenumbers $k$ and the corresponding growth rates. Their findings show that the amplitude of the dominant wavenumber does not grow exponentially over time, suggesting the instability is not in the linear regime predicted by the classical Taylor theory [1]. This difference in growth rate might be explained by the large density difference between the particles and air ($\rho_{particle} \cong 1000 \rho_{air}$) making particle inertia the dominating effect. Hence, it is conceivable that the RT instability theories developed for fluid-fluid systems maybe have limitations in describing RT-like instabilities in granular systems in which particle inertia dominates over the forces due to fluid-particle interactions.

Very recently, an additional RT-like instability was reported in particle-particle systems that are agitated by both vibration and a fluidizing gas [6]. Using two sets of particles with matching minimum fluidizing velocity $U_{mf}$ (but different diameters and densities) McLaren *et al.* [6] could ensure that both particle sets experience equal fluid-particle interaction forces. As $U_{mf}$ scales with $U_{mf} \propto d_i^{1/2} \rho_i$, where $d_i$ is the diameter of the particle and $\rho_i$ is the particle density, the lighter particles (subscript $i = L$) had a diameter of $d_L = 1.70$ mm and a density of $\rho_L = 2500$ kg/m$^3$, while the heavier particles (subscript $i = H$) had a diameter of $d_H = 1.105$ mm and a density of $\rho_H = 6000$ kg/m$^3$. The system was initialized by placing the heavy particles on top of a layer of light particles. Upon agitation by combined vibration and gas fluidization, the light particles penetrated upwards through the layer of denser particles exhibiting finger-like structures. Although the fingering-structures in the granular material mimic the structures that are commonly observed in conventional fluids, it was argued that the physics controlling the granular RT-like instability must be different involving a locally preferred gas channelling through the fingers of light particles and particle-particle contact forces pushing downwards the heavier particles. However, as McLaren *et al.* [6] have finely tuned the fluidization properties of the particle sets such that the fluid drag forces balance the weight of the particles and hence particle inertia is very likely to play a minor role, classic hydrodynamic RT instability theory might be applicable to such binary granular systems.

Therefore, the objective of this work is to conduct a quantitative analysis of the growth of fingers in a binary granular system allowing us to investigate its scaling behaviour and to assess whether it can be described by the classical RT theory developed for fluids. By complementing experiments with computational fluid dynamics coupled with the discrete element modelling (CFD-DEM) simulations we are able to probe the parameters that control the formation and growth of fingers in granular media and provide further insight into the underlying mechanism controlling this particular RT-like instability.

## 2. Methods

### A. Numerical Methodology

In the present work, CFD-DEM simulations were performed using the CFDEM®*Coupling* framework [21]. In the DEM part, particles are modelled as individual, freely moving entities with each particle $i$ having a certain mass $m_i$, velocity $\mathbf{u}_{pi}$ and angular velocity $w_i$. Forces acting on the particles lead to changes in their trajectories as described by Newton's second law of motion:

$$m_i \frac{d\mathbf{u}_{pi}}{dt} = \mathbf{f}_{ti} + \mathbf{f}_{fpi} + m_i \mathbf{g} , \quad (1)$$

and

$$I_i \frac{d\mathbf{w}_i}{dt} = \mathbf{T}_i . \quad (2)$$

where $\mathbf{f}_{ti}$ and $\mathbf{f}_{fpi}$ are the particle-particle contact force and the fluid-particle force, respectively, $\mathbf{g}$ is the gravitational acceleration, $I_i$ is the moment of inertia of particle $i$ and $T_i$ is the torque acting on particle $i$. The contact force between two contacting particles, $\mathbf{f}_{ti}$, is modelled via a Hertzian contact model using the following material properties: Youngs's modulus $E = 5$ MPa, Poisson ratio $v = 0.2$, coefficient of restitution $e = 0.30$, and inter-particle friction coefficient $\mu_p = 0.3$.

For the CFD part, the fluid phase is modelled by the locally averaged Navier-Stokes equations using a finite volume scheme:

$$\frac{\partial}{\partial t}(\varepsilon_f \rho_f) + \nabla \cdot (\varepsilon_f \rho_f \mathbf{u}_f) = 0 , \quad (3)$$

and

$$\frac{\partial}{\partial t}(\varepsilon_f \rho_f \mathbf{u}_f) + \nabla \cdot (\varepsilon_f \rho_f \mathbf{u}_f \mathbf{u}_f) = -\nabla p + \nabla \cdot (\varepsilon_f \boldsymbol{\tau}_f) \\ + \varepsilon_f \rho_f \mathbf{g} - \mathbf{F}_{fp} \quad (4)$$

where $\varepsilon_f$, $\rho_f$, $\mathbf{u}_f$, and $\boldsymbol{\tau}_f$ are the void fraction, the fluid density, the fluid velocity and the viscous stress of the fluid, respectively. The term $F_{fp}$ describes the momentum exchange between the particle and fluid phases and is given by:

$$\mathbf{F}_{fp} = \sum_{i=1}^{mc_i} \frac{1}{V_{mc}} \mathbf{f}_{fpi} , \quad (5)$$

where $mc$ is the index of a given fluid cell, $mc_i$ is the index of each particle in a given fluid cell $mc$ and $V_{mc}$ is the volume of the cell. The fluid force acting on each particle in the fluid cell is given as

$$\mathbf{F}_{fp} = -V_{pi} \nabla p + V_{pi} \nabla \cdot \boldsymbol{\tau}_f + \varepsilon_f \mathbf{f}_{di} \quad (6)$$

where $\varepsilon_f \mathbf{f}_{di}$ is the fluid drag acting on particle $i$. To model $\mathbf{f}_{di}$, the Koch-Hill correlation is used [22].

### B. Simulation setup

The simulation setup illustrated in Figure 1 mirrors the experimental setup reported by McLaren, *et al.* [6], i.e., using the same particle densities and sizes. The numerical simulations are initialized by placing a layer of heavier ($\rho_H$ = 6000 kg/m$^3$) particles of smaller diameter ($d_H$ = 1.16 mm) on top of lighter ($\rho_L$ = 2500 kg/m$^3$), but larger ($d_L$ = 1.70 mm) particles. For both types of particles some degree of polydispersity ($\pm 0.1 d_{H/L}$) is introduced to avoid crystallization. The width of the system $l_x$ = 200 mm is identical to the experimental setup of McLaren, *et al.* [6], while the filling height is reduced to $l_z$ = 200 mm for computational efficiency. The transverse thickness of the bed ($y$ direction) is 10 mm. The particle bed is agitated by a combination of an upward-directed gas flow (air) with an uniform inlet velocity $U$ = 1.13 m/s (density $\rho_{air}$ = 1.2 kg/m$^3$ and viscosity $\eta_{air}$ = 1.8·10$^{-5}$ Pa·s) and a vertical vibration with $z(t) = A\sin(2\pi f t)$, where $A$ = 1 mm and $f$ = 10 Hz.

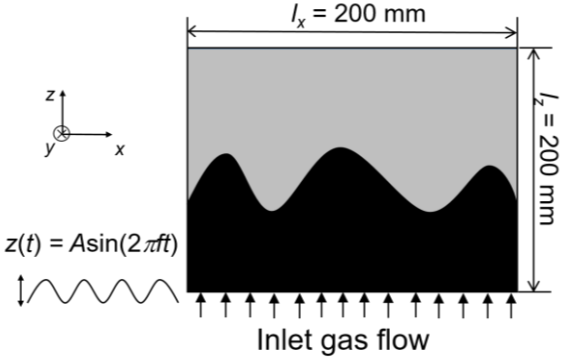

Figure 1: Illustration of the configuration of the simulated system. The gray color represents the heavy particles (diameter $d_H$ = 1.16 mm and density $\rho_H$ = 6000 kg/m$^3$) while the black colour denotes the light particles with $d_L$ = 1.70 mm and $\rho_L$ = 2500 kg/m$^3$. The container is vibrated vertically with $z(t) = A\sin(2\pi f t)$, where $A$ = 1 mm and $f$ = 10 Hz. A uniform upwards-directed gas flow (air) is injected at the bottom of the container with a velocity $U$ = 1.13 m/s (density $\rho_{air}$ = 1.2 kg/m$^3$ and viscosity $\eta_{air}$ = 1.8·10$^{-5}$ Pa·s).

The particle-wall interaction is modelled similarly to the particle-particle contacts (Hertzian contact model) whereby the wall is treated as a particle with an infinite radius. The pressure at the outlet (top) is fixed to 1.2·10$^5$ Pa and the size of a fluid cells is 5 × 5 × 5 mm$^3$). The DEM time step size is 10$^{-5}$ s, the CFD time step size is 5·10$^{-4}$ s and the CFD-DEM coupling interval is 5·10$^{-4}$ s.

### C. Experimental setup

The numerical simulations are complemented by experiments that are acquired in a setup identical to the one reported by McLaren *et al.* [6]. The particle bed is contained in a container made of acrylic glass with a width of 200 mm, height 500 mm and transverse thickness 10 mm (pseudo-2D bed). The experiment is initialized by filling the container first with a layer of light ($\rho_L$ = 2500 kg/m$^3$), large diameter particles ($d_L$ = 1.73±0.06 mm) followed by a layer of heavy ($\rho_H$ = 6000 kg/m$^3$), but smaller particles ($d_H$ = 1.17±0.06 mm). The thickness of each layer is 200 mm. The two types of particles are purchased from Sigmund Lindner GmbH and have the same properties as the particles used in the numerical simulations, i.e., a coefficient of restitution of 0.66 and a coefficient of friction of 0.4. During an experiment, humidified air is injected through a distributor placed at the bottom of the container (20 holes of diameter 1.1 mm) to fluidize the particles. The superficial air velocity was $U$ = 1.55 m/s (controlled via a Bronkhorst mass flow controller F-203AV). In addition, a vertical vibration is introduced via an electrodynamic shaker (Labworks Inc., ET-139) using a vibration strength of $\Gamma = A\omega^2/g$ = 0.45 with $\omega$ = 188 rad/s.

### D. Data analysis

In the experiments the interface between the two types of particles is recorded by high-speed camera imaging. Similarly, the predictions of the numerical simulations are visualized using the software ParaView to produce grey-scale images of the interface between the two types of particles. From these images the interface (red line in Figure 2) is identified by converting first the grey-scale image into a binary image by thresholding. The interface identification is insensitive to varying the thresholding value as doing so would only change individual pixels which in turn only affects high wavenumbers that are outside of the relevant range for granular fingering. At a given location in $x$, the interface position $z(x)$ is determined as the position in the $z$-direction at which the pixel value change from 1 (white) to 0 (black) [3].

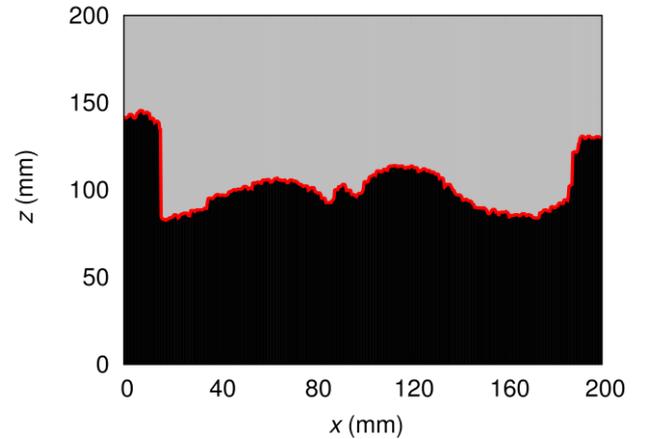

Figure 2: Snapshot of a numerical simulation at $\Delta t$ = 6.4 s. For this simulation a layer of heavy ($\rho_H$ = 6000 kg/m$^3$) particles with a small diameter ($d_H$ = 1.16 mm) is placed on top of lighter ($\rho_L$ = 2500 kg/m$^3$) but larger ($d_L$ = 1.70 mm) particles. The initial filling height for both layers is 100 mm. The system width is 200 mm and the transverse depth is 10 mm. The red curve denotes the interface between the heavy (gray) and light (black) particles identified through the algorithm proposed by ref [3].

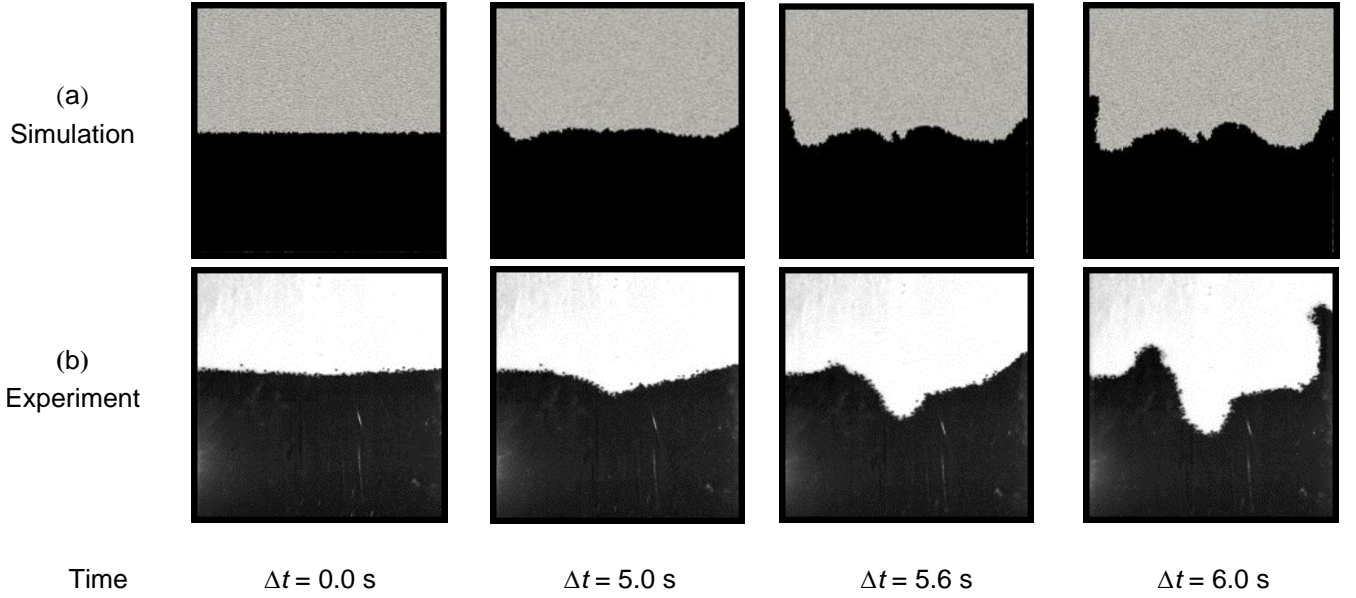

Time    Δt = 0.0 s    Δt = 5.0 s    Δt = 5.6 s    Δt = 6.0 s

Figure 3: Snapshots of the formation of granular fingers as observed in numerical simulations (a) and experiments (b) at various time steps $\Delta t$. In the numerical simulations the container is of width $l_x = 200$ mm with an initial filling height of $l_z = 200$ mm. The dimensions of the experimental setup are $l_x = 200$ mm and $l_z = 400$ mm. The transverse thickness in both the experiment and simulation is $l_y = 10$ mm. The heavier and smaller particles ($d_H = 1.16$ mm, $\rho_H = 6000$ kg/m$^3$) appear as grey (simulations) or white (experiments), while the lighter and larger particles ($d_L = 1.7$ mm, $\rho_L = 2500$ kg/m$^3$) are black.

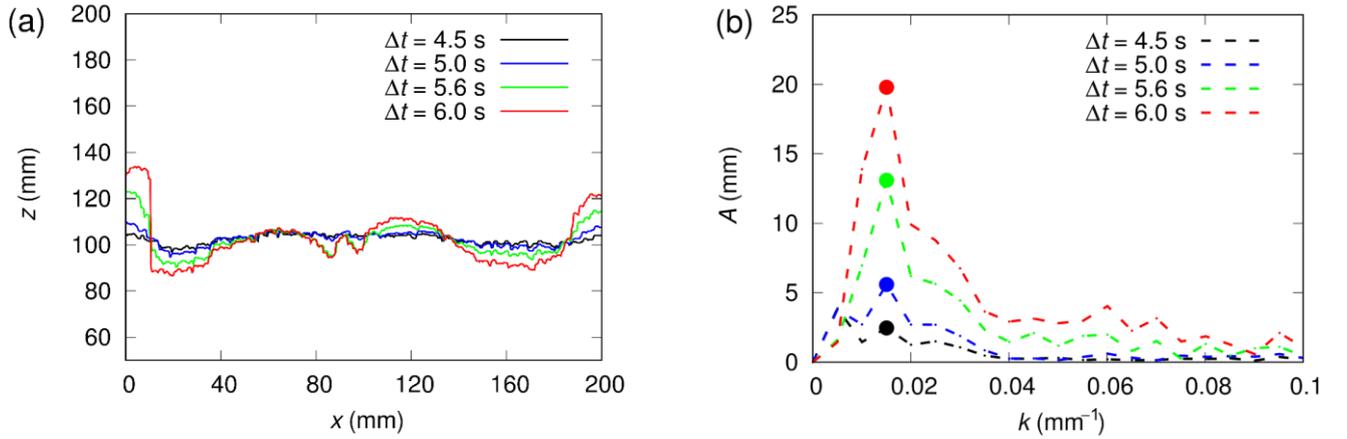

Figure 4: (a) The geometry of the interface between the light and heavy particles as a function of time as obtained from numerical simulations. (b) Fourier transform of the interface plotted in (a), where $k$ is the wavenumber, and $A$ is the corresponding amplititude. The wavenumber with the fastest growing amplitude, i.e. the dominant wavenumber is marked by solid circles. The dominant wavenumber ($k_c = 0.015$ mm$^{-1}$) is the characteristic wavenumber of the fingering instability.

The function $z(x)$ is subsequently Fourier transformed

$$Z(k) = \sum_{n=0}^{N-1} z(x_n) e^{-i2\pi nk/N}, \quad (7)$$

where $z(x_n)$ is the vertical position of the interface at the horizontal position $x_n$. The magnitude of the Fourier coefficient $Z(k)$ gives the amplitude of the $k$-th mode (wavenumber $k$). Typically, the 10 pixels next to the side walls are disregarded for the analysis.

## 3. Result and discussion

Figure 3 plots snapshots of the shape of the interface between the light and heavy particles in both the experiments and numerical simulations for certain times. Qualitatively the experimental observations and the numerical results agree well, however, a quantitative comparison between the experimental and numerical results from simple image analysis is not possible. Hence, for a quantitative comparison, we extract the function of the interface $z(x)$ (Figure 4(a)) and perform a Fourier transform. Figure 4(b) shows the amplitudes of the wavenumbers $k$ = 0.01–0.1 mm$^{-1}$ at selected time steps. By tracking the amplitude over time, we obtain a growth rate $n(k)$ that can be approximated by an exponential expression (vide infra). The characteristic (i.e. dominant) wavenumber $k_c$ of the fingering instability is the wavenumber with the fastest growing amplitude $A$. For the setup shown in Figure 4 with $k_c$ = 0.015 mm$^{-1}$ (characteristic wavelength $\lambda_c$ = $1/k_c$ = 66.7 mm), the characteristic wavenumber is coincidentally also the wavenumber with the highest amplitude $A$. In the binary granular system studied here, one wavenumber remains characteristic with time. Such a behaviour has been reported also for the classical fluid-fluid and liquid-particle systems [3], while for gas-particle systems the characteristic wavenumber is not necessarily constant with time, but instead was observed to generally increase with time (attributed to the branching of the fingers) [5].

Next, we study the growth rate of the fingering instability by plotting the amplitude of the characteristic wavenumber ($k_c$ = 0.015 mm$^{-1}$) as a function of time. Figure 5 plots both the results of the numerical simulations and the experiments (inset). Figure 5 shows that a sufficiently long time (~3 s) is required until an appreciable growth in amplitude occurs. Once sufficiently large perturbations have been formed, fingers emerge and grow with an exponential growth rate. Fitting an exponential function to this initial growth period, i.e. $A(k) = A_0 e^{(n(k)t)}$, yields growth rates of $n_E(k_c)$ = 1.75±0.04 s$^{-1}$ from the experimental data and $n_S(k_c)$ = 1.67±0.05 s$^{-1}$ from the numerical simulations, showing very good agreement between experiments and simulations. According to Taylor's theory [1], exponential growth only occurs for very small perturbations ($A < 0.4\lambda_c$). Therefore, only the growth period in which A $< 0.4\lambda_c$ was considered for the fitting, i.e. $\Delta t$ = 3–6 s.

### A. Rayleigh-Taylor instability theory

Our experimental and numerical modelling data have shown that there are some striking similarities with regards to the finger morphology and growth rate between the fingering instability in binary granular systems and their classic fluid counterpart. In the following we investigate if the dispersion relation between the wavenumber $k$ and its respective growth rate $n(k)$ as derived for conventional fluid-fluid systems is equally applicable to binary granular systems. The dispersion relation is a result of the linear RT instability theory which was first reported by Harrison [23] and subsequently developed further by Chardrasekhar [13], and Bellman and Pennington [24]. The linear RT instability theory models the growing perturbations of an initially cosine shaped, unstable interface between a denser fluid on top of a less dense fluid using hydrodynamic arguments. To derive the dispersion relation for a binary granular system, it is assumed that both granular media are viscous and incompressible and surface tension is neglected. The full derivation of the implicit dispersion relation for a binary granular system, which is based on the work by Bellman and Pennington [24], is given in Appendix A. Here only the final form of the dispersion relation is given as:

$$\left[-g\phi(\rho_H - \rho_L)K + \phi(\rho_L + \rho_H)n(K)^2\right]M + 4n(K)K = 0 \quad (8)$$

where $M$ is given by

$$M = \frac{1}{\eta_H K + (\eta_L^2 K^2 + \phi\rho_L n(K)\eta_L)^{1/2}} + \frac{1}{\eta_L K + (\eta_H^2 K^2 + \phi\rho_H n(K)\eta_H)^{1/2}} \quad (9)$$

Here $K$ is the angular wavenumber, $K = 2\pi k$, and $\phi$ is the bulk solid fraction. $\eta_L$ and $\eta_H$ are the viscosity of the light and heavy granular medium, respectively. However, the viscosity of granular materials is not a constant, but dependents on the local state of the system (i.e., it is controlled by the local pressure, solid fraction and shear rate [25,26]). Since the characteristic wavelength of the granular system at hand is known, i.e., $\lambda_c$ = 66.7 mm, we can derive an explicit equation for the granular viscosity from the dispersion relation when assuming that both granular media have the same viscosity, i.e., $\eta_L = \eta_H$. To derive an equation for the viscosity, Bellman and Pennington [24] propose to simplify Eq. (8) by assuming that the growth rates of high wavenumbers are limited by viscosity and the growth rate decreases with increasing wavenumber, i.e. $\rho n(k) \ll \eta k^2$. This simplification yields:

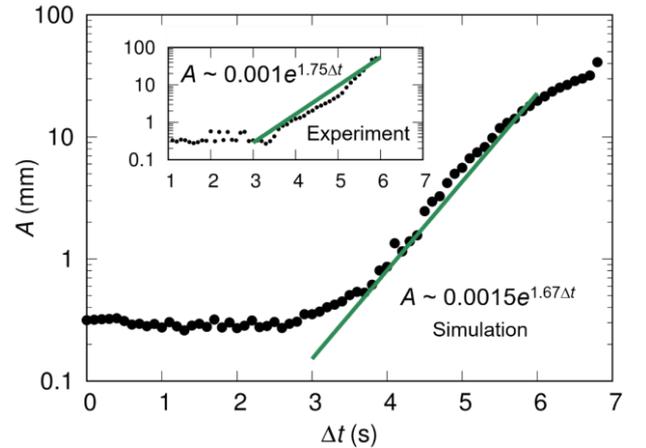

Figure 5: Growth of the amplitude $A$ of the characteristic wavenumber $k_c$ = 0.015 mm$^{-1}$ as a function of time as obtained from numerical modelling and experiments (inset). The insert figure has the same axis labels as the main figure. The green lines are exponential fits during the initial growth period.

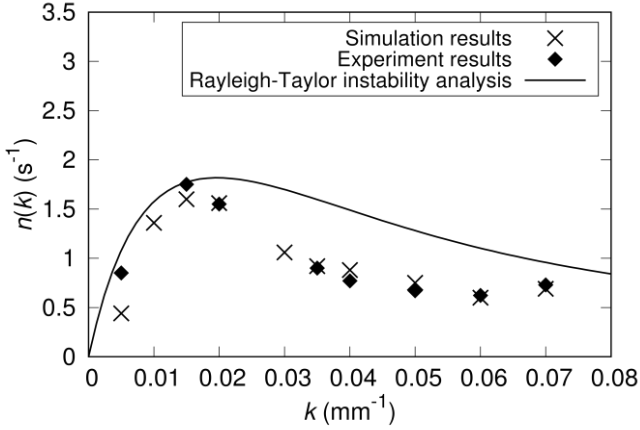

Figure 6: The dispersion relation, i.e., the growth rate $n(k)$ as a function of the wavenumber $k$ as predicted by RT theory [Eq. (8)] (solid line) and the respective data obtained from numerical simulations (×) and experiments (◆).

$$n^2(K) + 2\frac{\eta_H + \eta_L}{\phi\rho_H + \phi\rho_L}K^2 n(K) - gK\frac{\phi\rho_H - \phi\rho_L}{\phi\rho_H + \phi\rho_L} = 0. \quad (10)$$

Since the amplitude of the characteristic wavenumber has the highest growth rate, the characteristic wavenumber can be found by differentiating Eq. (10) with respect to $K$ and setting $dn(K)/dK = 0$:

$$K_c = \frac{(\phi\rho_H + \phi\rho_L)g}{4n(K_c)(\eta_H + \eta_L)}. \quad (11)$$

Substituting Eq. (11) into Eq. (10), the characteristic wavelength $\lambda_c = 2\pi/K_c$ is given as [27]:

$$\lambda_c = 2\pi\left(\frac{8(\eta_H + \eta_L)^2}{\phi^2(\rho_H^2 - \rho_L^2)g}\right)^{1/3}. \quad (12)$$

To calculate the granular viscosity from Eq. (12), the bulk solid fraction $\phi$ is required. The bulk solid fraction can be obtained from the numerical simulations via coarse-graining (described in detail in Appendix B). Coarse-graining yields $\phi = 0.6 \pm 0.03$. Using $\phi = 0.6$ in Eq. (12), we obtain a granular viscosity of $\eta = 1.95$ Pa·s. This value is close to the value of 1.25 Pa·s, that is predicted by a viscosity model for sheared and fluidized granular systems [26], i.e. using the correlation $\eta_{eff} = \eta_{air}\exp(k_0 + k_1\dot{\gamma} + k_2\phi)$, where $k_0 = 2$, $k_1 = -1.78$, $k_2 = 20.20$, $\eta_{air} = 1.8 \cdot 10^{-5}$ Pa·s and approximating the shear rate by the growth rate of the characteristic wavenumber, i.e. $n(k) = 1.67$ s$^{-1}$.

Having estimated now the granular viscosity we can plot the dispersion relationship between $n(k)$ and $k$ for the binary granular system studied here by numerically solving Eq. (8). The results are given in Figure 6 and we observe that classical RT theory predicts generally very well the growth rate of the interface amplitude $A(k)$ as a function of the wavenumber $k$ (using both experimental and numerical data), although there is some overprediction of the numerical and experimental data, for higher wavenumbers (possibly due to the fact that the shear rate was approximated by the growth rate of the characteristic wavenumber).

## B. Role of system size

Figure 3 shows that fingers appear typically at both side walls, probably due to relatively large perturbations of the granular packing near the side walls (i.e. large gradients in solid fraction in the vicinity of the side walls). This observation triggers the question of the effect of system size on the dynamics of granular fingering. To address this question, we have simulated systems of increasing widths, i.e. with $l_x = 200$, 400 and 600 mm, while keeping the filling height constant at $l_z = 200$ mm. The characteristic wavelengths and corresponding growth rates obtained in these three systems are plotted in Figure 7. For the smallest system, i.e. $l_x = 200$ mm, the characteristic wavelength is lower, viz. $\lambda_c = 66.7$ mm than for the two wider systems. However, the characteristic wavelength reaches an asymptotic value of $\lambda_c \sim 100$ mm for $l_x \geq 400$ mm. Inversely, the growth rate of the characteristic wavelength is higher for the smallest system size, i.e. $n(k_c) = 1.67$ s$^{-1}$ for $l_x = 200$ mm, but again reaches an asymptotic value of $n(k_c) = 1.45$ s$^{-1}$ for $l_x \geq 400$ mm. These results suggest that for systems with a width $l_x \geq 400$ mm the characteristic wavelength and growth rates become independent of system size.

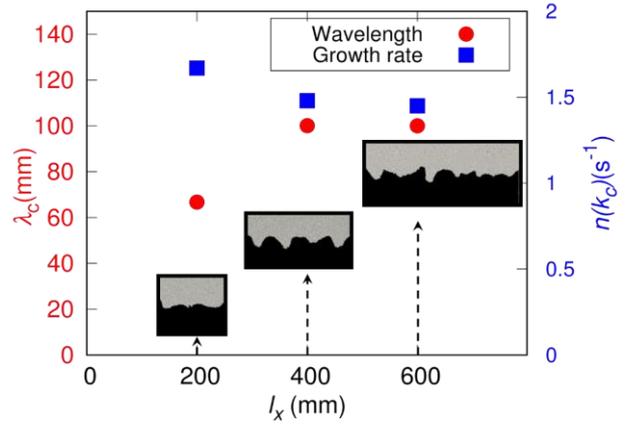

Figure 7: The characteristic wavelength (●) and its growth rate (■) as a function of system width $l_x$.

## C. Role of inter-particle friction

Previous research has suggested that particle friction, through its influence on the effective viscosity of the granular medium, affects the formation and dynamics of granular fingering [6,27,28]. To probe its influence in more detail, we performed additional simulations with varying coefficients of particle friction $\mu_p$ and coefficients of restitution $e$. In these simulations the system width was set to $l_x = 400$ mm to eliminate any effect of the system size, while maintaining the computational efficiency of relatively small systems.

Figure 8(a) plots the initial growth rate of the characteristic wavenumber as a function of $\mu_p$. For $\mu_p \geq 0.1$ the growth rate of the characteristic wavenumber reaches an asymptotic value of $n(k_c) = 1.45$ s$^{-1}$, while for lower coefficients of friction the growth rate of the characteristic wavenumber increases strongly, reaching $n(k_c) = 5.3$ s$^{-1}$ for $\mu_p = 0$. An increase of the growth rate of the characteristic wavenumber with decreasing $\mu_p$ is predicted also by classic RT theory (effective viscosity of dense granular media decrease with decreasing $\mu_p$ [28]). However, as plotted in Figure 8(b) the magnitude of the growth rate of the characteristic wavenumber for $\mu_p \to 0$ ($n(k_c) = 5.3$ s$^{-1}$) is very high; indeed for $k \leq 0.02$ mm$^{-1}$ ($k_c = 0.005$ mm$^{-1}$) the growth rates in a frictionless granular medium as determined from numerical simulations (◆ in Figured 8(b)), exceed the upper bound of the dispersion relation using classic, non-viscous RT theory (dashed line). On the other hand, the prediction of the classic RT theory (dash-dotted line) for a system with $\mu_p = 0.3$ ($\eta_{eff} = 1.95$ Pa·s) agrees very well with the results of the numerical simulations (● in Figured 8(b)). Based on our results of granular systems with varying coefficients of friction it appears that granular systems with $\mu_p \leq 0.1$ are not predicted well by classic RT theory. The different behaviour of such systems is visualized in more detail in Figure 9(a–d). In Figure 9(a), i.e. $\mu_p = 0$, we observe spikes at the fingers, while such spikes are absent in both classic RT fingers and in granular fingers for $\mu_p \geq 0.1$ (Figure 9(d)). Such spike features, as observed for $\mu_p = 0$, resemble structures that have been observed in the hydrodynamic Richtmyer-Meshkov (RM) instability [29]. The RM instability occurs when a light fluid is accelerated into a heavy fluid by a shockwave or impulsive flow, which amplifies any perturbation of the interface due to the reflection of the shock wave at the interface.

Although the granular system that is studied here is not driven by impulsive flow, for the case $\mu = 0$ we observe the upwards motion of void bands through the interface that could generate shock-like effects (Figure 9(e)). Importantly, the formation of such void bands is largely suppressed for $\mu_p \geq 0.1$ (see Appendix C), since the injected vibrational energy is dissipated quickly through inter-particle friction. For $\mu_p < 0.1$ void bands form at the bottom of the system and move upwards. Somewhat surprisingly, at the location of a finger, the void band is interrupted, while the void band exists at the left- and right-hand side of the finger ($\Delta t = 1.0$ s in Figure 9(e)). It appears that the spike-like features at the fingers arise due to particles being lifted up by the passing void band. Owing to the absence of friction, the finger is also growing faster in the lateral dimension compared to the dynamics of the fingers in the frictional cases (Figure 9(b–d)). Although there is some geometric similarity between the frictionless case and structures observed in RM instabilities, fingers still show an exponential growth rate in the frictionless case, while a linear growth rate would be expected for the RM instability. Interestingly, also for a gas-particle RM instability in which an explosive gas penetrates into a radial, granular Hele-Shaw cell, an exponential finger growth has been reported [30] and it was also argued that high voidage due to dilation plays a key role in the observed RM instability.

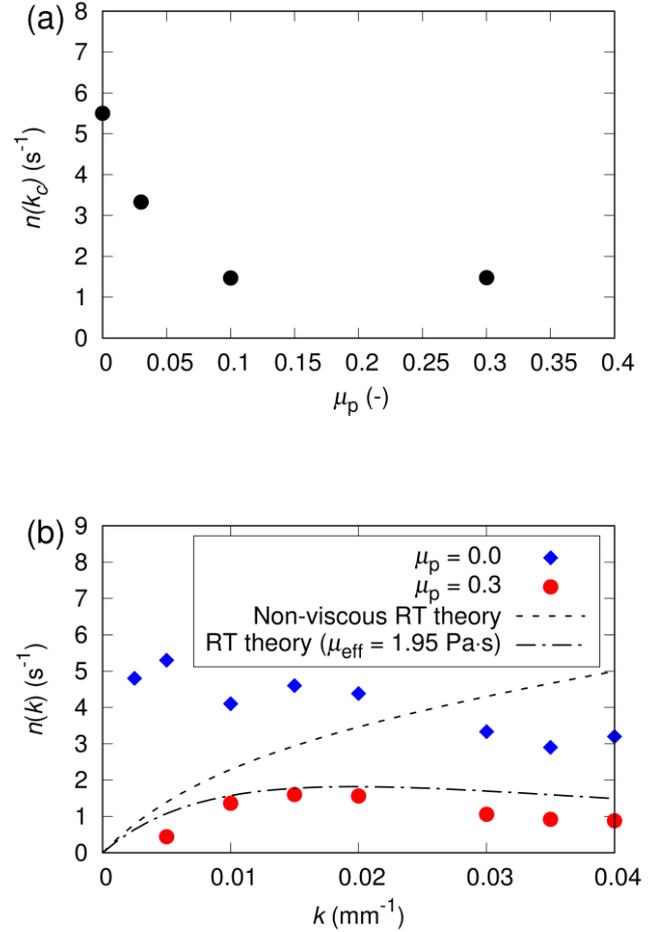

Figure 8: (a) Growth rate $n(k_c)$ of the characteristic wavenumber as a function of the coefficient of friction $\mu_p$. (b) Dispersion relation between the growth rate, $n(k)$ and the wavenumber $k$. The markers plot data obtained from numerical simulations for $\mu_p = 0.0$ and $\mu_p = 0.3$. The corresponding granular viscosity for $\mu_p = 0.3$ is $\eta = 1.95$ Pa·s (using Eq. (12)). (– · –) Prediction of the classic RT theory for $\eta_{eff} = 1.95$ Pa·s. For $\mu_p = 0.0$ the system does not exhibit any RT-like instability, therefore Eq. (12) cannot be used to estimate the granular viscosity. However, classic RT theory for non-viscous fluids ($\eta_{eff} = 0$ Pa·s) gives a theoretical upper bound for $n(k)$ which is given by (- - -).

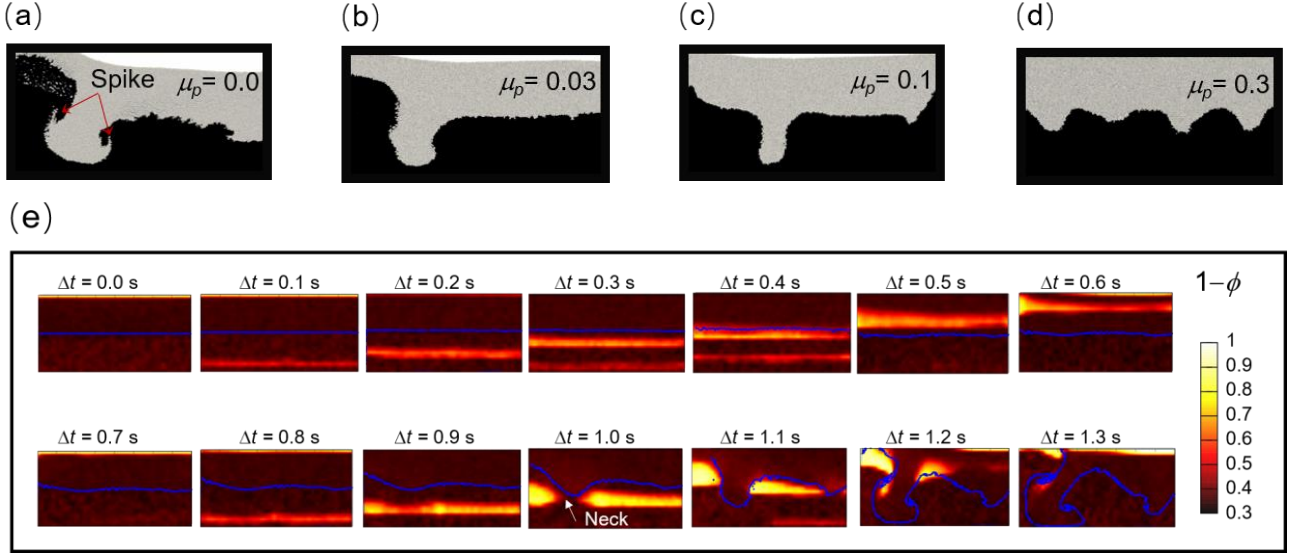

Figure 9: (a), (b), (c), (d) Shape of the interface for varying coefficients of friction $\mu_p$. (e) Time series visualizing the void fraction $(1-\phi)$ in a layered granular system using $\mu_p = 0$. The size of the numerical domain was $l_x \times l_y \times l_z = 400 \times 10 \times 200$ mm$^3$. The blue line denotes the interface between the heavy and light particle layers.

Unlike the coefficient of friction, the coefficient of restitution $e_p$ is found to have a negligible effect on the dynamics of granular fingering. Simulating three granular systems with varying $e_p$, i.e. $e_p = 0.06, 0.30, 0.98$, we observe very similar initial growth dynamics as shown in Figure 10.

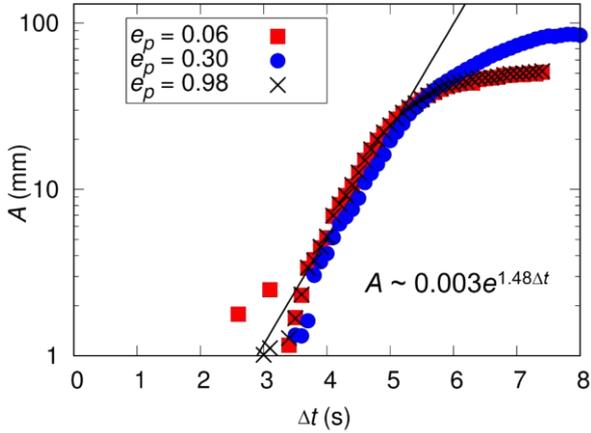

Figure 10: Influence of the coefficient of restitution $e_p$ on the growth rate of the characteristic wavenumber. The solid line plots a (fitted) growth rate of 1.48 s$^{-1}$.

### D. Yield criterion for granular RT instability

In the previous section it was discussed how the growth of perturbations at the interface of two layers of granular materials of different densities depend on inter-particle friction. However, Figures 3–5 also show that there is a considerable time lag between the start of the agitation of the system and finger formation and growth to occur, suggesting that an initial resistance, i.e. a yield criterion, has to be overcome for fingers to form. The granular system considered here falls into the category of a dense granular system/flow owing to its high solid fraction of ~0.6. Similar to toothpaste, foam, etc., dense granular flows exhibit a viscoplastic-like behaviour, i.e. there exists a critical shear stress below which flow is not maintained and the rheology depends on the shear rate. A model that aims to describe the complex rheology of dense granular systems is the $\mu(I)$ model [28], whereby the inertial number $I$ is a function of the particle size $d$, the particle density $\rho_p$, the shear rate $\dot{\gamma}$ and pressure $P$, viz. [31]:

$$I = \frac{\dot{\gamma} d}{\sqrt{P/\rho_p}}. \qquad (13)$$

The shear stress $\tau$ is expressed as a function of $I$ as

$$\tau = \mu(I) P \qquad (14)$$

where $\mu(I)$ is bulk friction (different from the inter-particle friction) and given through an empirical friction law as [28,32]:

$$\mu(I) = \mu_s + (\mu_2 - \mu_s)/(I_0/I + 1) \qquad (15)$$

Here, $\mu_s$ is a critical friction coefficient at zero shear rate which depends on the particle properties such as the coefficient of inter-particle friction and particle shape. The critical friction is $\mu_s \approx 0.38$ for spherical particle systems, based simulation and experimental results [33]. In the following, $\mu_s = 0.38$ will be used for simplicity. The $\mu(I)$-rheology model is illustrated in Figure 11 and compared to a typical viscoplastic rheology model for fluids (see insert in Figure 11) which is described by:

$$\tau = \tau_y + \tau_s \dot{\gamma}^n (\dot{\gamma}/\dot{\gamma}_s)^n \qquad (16)$$

where $\tau_y$ is the yield stress and $n$ is the flow index. Similar to the yield stress in a viscoplastic fluid (dashed line in the

insert in Figure 11), the yield stress for a dense granular system is given by:

$$\tau_y = \mu_s P \quad (17)$$

Hence, the yield stress originates from the granular pressure $P$ and is not an intrinsic property of a granular material [34]. Furthermore, the yield criterion, Eq. (17), describes a yield transition from a static system. However, the present system is not static as the individual particles are agitated by vertical vibration and a fluidizing gas flow, yet the relative motion between heavy and light particles before fingering occurs is small, allowing to make the assumption of a pseudo-static system. Furthermore, the following discussion is limited to $\mu_p \geq 0.1$ such that void bands are absent.

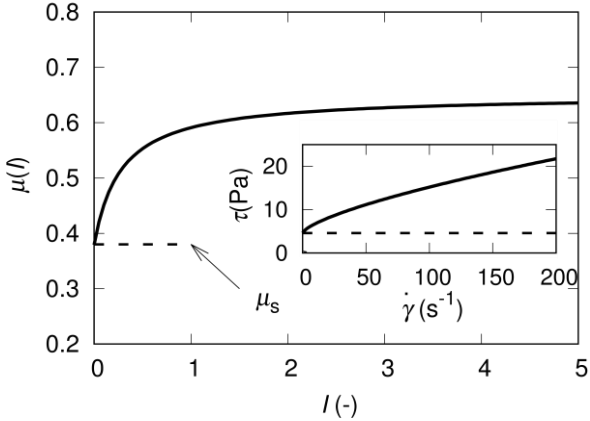

Figure 11: $\mu(I)$ local rheology model for a dense granular system given by Eq. (15); Insert: Herschel-Bulkley rheology model for an oil-based drilling fluid, i.e. a typical viscoplastic fluid [35].

As buoyancy is the driving force for the granular RT instability, a yield parameter $Y$ can be established as:

$$Y = \frac{(\rho_H - \rho_L)\phi g \lambda_c}{\tau_y}, \quad (18)$$

where $\lambda_c$ is the characteristic wavelength of the granular RT instability, $\phi$ is the solid fraction of the bulk and $\rho_H$ and $\rho_L$ are densities of the heavy and light particle, respectively. The yield stress $\tau_y$ is given by Eq. (17), which requires information on the granular pressure $P$ at the interface. The granular pressure at the interface can be determined from the inter-particle contact forces using coarse-graining (described in Appendix B). The pressure profile as a function of the vertical position $z$ is initially hydrostatic, but very rapidly a pressure profile develops that leads to an instability at the interface, i.e. the higher granular pressure in the heavier particle layer pushes down onto a lower granular pressure in the layer containing the lighter particles (see Appendix D). Taking the granular pressure at the interface, i.e. $P(z = 100 \text{ mm})$, we can calculate the yield parameter $Y$ as given in Eq. (18). Figure 12 plots $Y$ over time for systems of varying $l_x$ and $e_p$, while the coefficient of inter-particle friction was fixed to $\mu_p = 0.3$. To determine the onset of exponential finger growth we calculate the intersection of a horizontal line fitted to the initial period where the amplitude of the characteristic wavenumber does not change (see Figure 5) and the exponential fit to the growth period (green line in Figure 5). In Figure 12 black symbols denote points in time in which no exponential finger growth was observed, while red symbols denote exponential finger growth. The results plotted in Figure 12 suggest a critical yield parameter of $Y_C = 15.7$ for RT-like finger instabilities to occur. Interestingly, this critical value of $Y_C = 15.7$ obtained here is very close to the value of $Y_C = 15\pm3.6$ that has been established for thermal plumes in the Earth's lower mantle to occur (which can also be classified as a visco-plastic material) [7,36]. Similar to Eq. (18), also the yield parameter for a thermal plume is defined as the ratio of the buoyancy driven shear stress to the yield stress. Although the apparent similarities between the finger instabilities in granular materials as observed in this work and the rise of thermal plumes in the Earth's lower mantle suggest that the critical yield parameter value of $Y_C = 15$ might be a more general scaling constant for the occurrence of fingering phenomena in viscoplastic fluids, further research is necessary to validate this hypothesis.

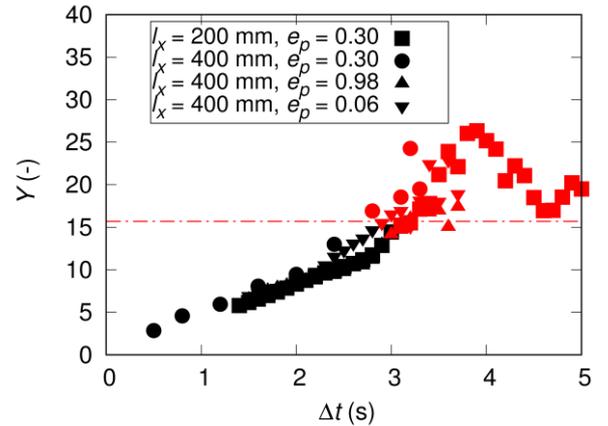

Figure 12: The yield parameter $Y$ as a function of time. The onset of exponential finger growth is calculated as the intersection of a horizontal line fitted to the initial period when the amplitude of the characteristic wavenumber does not vary with time with an exponential fit to the growth period (see Figure 5). Black colour symbols indicate $\Delta t$ for which no finger grow is observed, while red symbols denote $\Delta t$ in which RT-like, exponential finger growth proceeds. The red dash-dotted line gives the critical yield parameter $Y_C = 15.7\pm1.98$. The data plotted are obtained from a series of systems with varying widths, $l_x$, and varying coefficients of restitution, $e_p$: (■) $l_x = 200$ mm, $e_p = 0.30$; (●) $l_x = 400$ mm, $e_p = 0.30$; (▼) $l_x = 400$ mm, $e_p = 0.98$; (▲) $l_x = 400$ mm, $e_p = 0.06$; (●) $l_x = 400$ mm, $e_p = 0.3$. The coefficient of friction was fixed to $\mu_p = 0.3$ in all simulations.

## 4. Conclusion

The present work investigates binary granular systems in which a granular medium of dense and small particles is layered on top of a granular medium of light and large particles. Finger instabilities akin to the hydrodynamic Rayleigh-Taylor (RT) instability, emerge at the interface between the two granular media when agitated by a combination of vertical vibration and a fluidizing gas flow [6]. The results presented here confirm that classic RT instability theory can be used to describe the behaviour of fingering in dissipative binary granular systems (i.e. inter-particle friction coefficient $\mu_p \geq 0.1$) as:

- The characteristic wavenumber is constant over time.
- The initial growth rate of the characteristic wavenumber is exponential.
- The dispersion relation for the growth rate $n(k)$ as a function of the wavenumber $k$ follows a very similar behaviour in the fluid and granular systems.

For $\mu_p < 0.1$ the system behaviour changes leading to the formation of spike-like features in the fingers which do not resemble a typical RT behaviour, but instead show some similarity to a Richtmyer-Meshkov-type instability. Our results also suggest that by treating the binary granular material ($\mu_p \geq 0.1$) as a viscoplastic material, we can define a yield criterion $Y$ to predict the onset of fingering. If the yield criterion is below a critical value of $Y_c \approx 15$ fingering is not observed, while for $Y > 15$ fingers emerge and grow exponentially. A critical value of $Y_c \approx 15$ has also been found for other viscoplastic materials, such as the Earth's lower mantle in which thermal plumes rise for $Y > 15$ [7,36]. This suggests that $Y_c = 15$ might be a general scaling constant for the emergence of instabilities in viscoplastic materials. However, further research is required to confirm this hypothesis.

# Acknowledgment

We are thankful to the Swiss National Science Foundation (Grant No. 200020_182692) and the China Scholarship Council (M.L.) for partial financial support of this work. M. L acknowledges Prof. C. Boyce for his introduction to the CFD-DEM simulation techniques. This publication was created as part of NCCR Catalysis (grant number 180544), a National Centre of Competence in Research funded by the Swiss National Science Foundation.

# Appendix A: Derivation of the dispersion relation

To derive an equation for the dispersion relation in binary granular systems we follow the approach of Bellman and Pennington [24]. As the transverse thickness ($y$ direction) is very small compared to the other two dimensions, the system can be considered 2D with a heavy granular medium on top of a light granular medium. The initial unperturbed interface is located at $z = 0$. In addition, surface tension can be neglected in granular systems. The goal is to find an equation describing the location of the interface $z(x,t)$ when an initial perturbation of $\cos(kx)$ is applied. The governing equations of the system containing two viscous and incompressible granular media are given by,

$$\frac{\partial u}{\partial x} + \frac{\partial v}{\partial z} = 0, \tag{A1}$$

$$\frac{\partial u}{\partial t} = -\frac{1}{\rho}\frac{\partial p}{\partial x} + \frac{\eta}{\rho}\nabla^2 u, \tag{A2}$$

$$\frac{\partial v}{\partial t} = -\frac{1}{\rho}\frac{\partial p}{\partial z} - g + \frac{\eta}{\rho}\nabla^2 v, \tag{A3}$$

where $\eta$ is the dynamic viscosity, $\rho$ is the density, and $u$ and $v$ are the velocity in the horizontal and vertical direction, respectively. By introducing the potential functions $\Theta$ and $\Psi$, the solutions for Eq. (A1–A3) can be obtained as:

$$u = -\frac{\partial \Theta}{\partial x} - \frac{\partial \psi}{\partial z}, \tag{A4}$$

and

$$v = \frac{\partial \psi}{\partial x} - \frac{\partial \Theta}{\partial z}. \tag{A5}$$

The pressure is given by the Bernoulli equation,

$$p = p_0 - gz(x,t) - \frac{\partial \Theta}{\partial t}, \tag{A6}$$

where $p_0$ is the mean pressure at the (unperturbed) interface. Substituting Eq. (A4) and (A5) into Eq. (A1–A3) yields:

$$\Delta \Theta = 0, \tag{A7}$$

$$\frac{\partial \Psi}{\partial t} = \frac{\eta}{\rho}\Delta\Psi. \tag{A8}$$

Solutions for the potential functions $\Theta$ and $\Psi$ which satisfy (A7) and (A8) are given by Lamb [37]; the solutions for the heavy and light granular medium are denoted by subscript $H$ and $L$, viz.:

$$\begin{bmatrix} \Theta_H \\ \Theta_L \end{bmatrix} = \begin{bmatrix} Ae^{-Kz+nt}\cos(Kx) \\ Ce^{Kz+nt}\cos(Kx) \end{bmatrix}, \tag{A9}$$

$$\begin{bmatrix} \psi_H \\ \psi_L \end{bmatrix} = \begin{bmatrix} Be^{-m_H z+nt}\sin(Kx) \\ De^{-m_L z+nt}\sin(Kx) \end{bmatrix}, \tag{A10}$$

where $m_H^2 = K^2 + \frac{n\phi\rho_H}{\eta_H}$, $m_L^2 = K^2 + \frac{n\phi\rho_L}{\eta_L}$ and $n$ is the growth rate of a given wavenumber $k$ ($K = 2\pi k$). The pressure in both granular media is:

$$\begin{bmatrix} p_H \\ p_L \end{bmatrix} = \begin{bmatrix} p_0 - gz(x,t)\phi\rho_H - \phi\rho_H \frac{\partial \Theta_H}{\partial t} \\ p_0 - gz(x,t)\phi\rho_L - \phi\rho_L \frac{\partial \Theta_L}{\partial t} \end{bmatrix}. \tag{A11}$$

The boundary conditions at the interface between the two media are:

$$u_H = u_L, \ v_H = v_L \tag{A12}$$

$$-p_H + 2\eta_H \frac{\partial v_H}{\partial z} = -p_L + 2\eta_L \frac{\partial v_L}{\partial z}, \quad (A13)$$

$$\eta_H(\frac{\partial v_H}{\partial x} + \frac{\partial u_H}{\partial z}) = \eta_L(\frac{\partial v_L}{\partial x} + \frac{\partial u_L}{\partial z}), \quad (A14)$$

The vertical velocity of the interface $v$ is obtained by taking the partial differential of $z(x,t)$ with respect to $t$ yielding:

$$v - \frac{\partial z}{\partial x}u - \frac{\partial z}{\partial t} = 0. \quad (A15)$$

Assuming that the initial perturbation has a small amplitude compared to its wavelength, the nonlinear term $(\partial z/\partial x)u$ in Eq. (A15) can be neglected:

$$\frac{\partial z}{\partial t} = v. \quad (A16)$$

Since $v = v_H = v_L$, see Eq. (A12), we can calculate $v$ by substituting Eq. (A9) and Eq. (A10) into Eq. (A5) and integrate over $t$ to obtain $z(x, t)$ as:

$$z = K(A+B)n^{-1}e^{nt}\cos Kx. \quad (A17)$$

In this equation the constants $A$ and $B$ are still unknown. The constants $A$ and $B$ (and $C$, $D$) are determined by substituting Eq. (A9) and (A10) into Eq. (A4) and (A5) and the resulting equation into the boundary conditions given by Eq. (A12–A14). This yields a system of four equations that are linear in $A$, $B$, $C$ and $D$ (curious readers find the system in Bellman and Pennington [24] as equation (2.20)). A non-trivial solution for the system of equations exists if the determinant of the coefficient matrix of the four linear equations is zero, which yields the implicit dispersion relation:

$$\left[-\beta + \phi(\rho_H + \rho_L)n^2\right]\left[(\eta_H K + \eta_L m_L) + (\eta_L K + \eta_H m_H)\right] +$$
$$4nK(\eta_H K + \eta_L m_L)(\eta_L K + \eta_H m_H) = 0. \quad (A18)$$

If $\beta$ is positive, there will be at least one root of $n$ that has a positive real part, i.e., the interface is unstable and any perturbation of the interface grows in amplitude. As expected, this is the case if $\rho_H > \rho_L$.

In a final step we rewrite the implicit dispersion relation given by Eq. (A18) using the definitions for $m_H$ and $m_L$ to obtain:

$$\left[-g\phi(\rho_H - \rho_L)K + \phi(\rho_L + \rho_H)n^2\right]M + 4nK = 0, \quad (A19)$$

where $M$ is

$$M = \frac{1}{\eta_H K + (\eta_L^2 K^2 + \phi\rho_L n\eta_L)^{1/2}} + \frac{1}{\eta_L K + (\eta_H^2 K^2 + \phi\rho_H n\eta_H)^{1/2}}. \quad (A20)$$

# Appendix B: Coarse-graining method

Here we use coarse-graining (CG) as described in [38] to obtain the granular pressure. To this end, we define a CG volume, which is bounded by a sphere located at **r** with radius $w$. The CG density at **r** is defined as,

$$\rho(\mathbf{r},t) = \sum_{i=1}^{N} m_i G(\mathbf{r} - \mathbf{r}_i(t)), \quad (B1)$$

where the sum is taken over the particles located in the CG volume and $G(\mathbf{r}-\mathbf{r}_i(t))$ is the CG function. Here, we have chosen the Heaviside function $G(\mathbf{r})=H(w-\|\mathbf{r}\|)/V$ as the CG function, where $V$ is the volume of the CG sphere. The solid fraction is calculated as

$$\phi = \rho(\mathbf{r},t)/\rho_p, \quad (B2)$$

where $\rho_p$ is the particle density.

The CG moment density is given as

$$\rho(\mathbf{r},t)\mathbf{u}(\mathbf{r},t) = \sum_{i=1}^{N} m_i G(\mathbf{r} - \mathbf{r}_i(t)) \cdot \mathbf{u}(\mathbf{r}_i,t), \quad (B3)$$

and the CG velocity is calculated via

$$\mathbf{u}(\mathbf{r},t) = \sum_{i=1}^{N} m_i G(\mathbf{r} - \mathbf{r}_i(t)) \cdot \mathbf{u}(\mathbf{r}_i,t)/\rho(\mathbf{r},t). \quad (B4)$$

The CG stress tensor is derived based on the momentum conservation equation as:

$$\sigma_{\alpha\beta}(\mathbf{r},t) = -\frac{1}{2}\sum_{i,j}\mathbf{f}_\alpha^{i,j}(t)\mathbf{r}_\beta^{i,j}\int_0^1 ds\left(G(\mathbf{r} - \mathbf{r}_j(t) + s\mathbf{r}_{ij}(t))\right)$$
$$-\sum_i m_i u_\alpha^{i'}(t)u_\beta^{i'}(t)G(\mathbf{r} - \mathbf{r}_i(t)), \quad (B5)$$

where $\alpha$, $\beta$ denote the Cartesian coordinates $x$, $y$, $z$, $\mathbf{f}_\alpha^{i,j}$ is the $\alpha$-th component of the inter-particle contact force of the contacting particles $i$ and $j$, $\mathbf{r}_\beta^{i,j}$ is the vector that points from the position of particle $i$ to particle $j$ and $u_\alpha^{i'}$ is the fluctuation of the velocity of particle $i$ with respect to the CG velocity $u(\mathbf{r},t)$. The granular pressure is given by the trace of the stress tensor, i.e. $P = 1/3(\sigma_{xx} + \sigma_{yy} + \sigma_{zz})$.

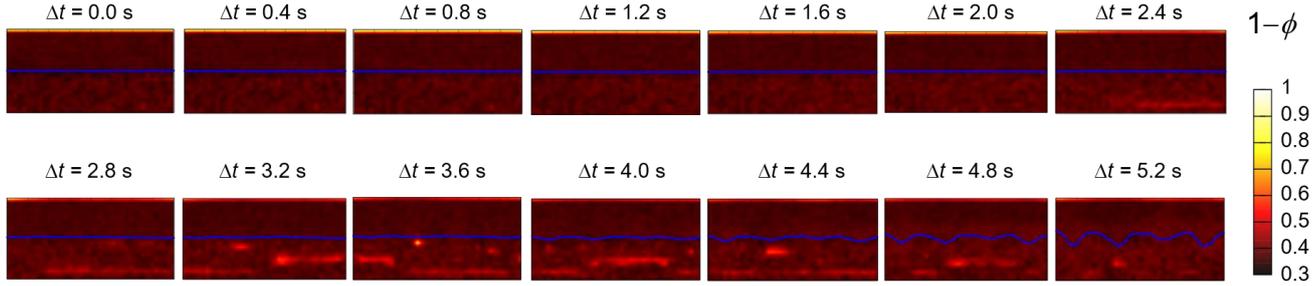

Figure C1: Time series of the void fraction $(1-\phi)$ in a numerical system using an inter-particle friction coefficient of $\mu_p = 0.3$. The blue line denotes the interface between the heavy and light granular medium.

# Appendix C: Void fraction for $\mu_p \geq 0.1$

In Figure 10(d) we observe the rise of void bands through a binary granular system when using a low inter-particle friction coefficient, i.e. $\mu_p < 0.1$. Formation of these void bands is largely supressed for $\mu_p \geq 0.1$ which is visualized in Figure C1.

# Appendix D: Granular Pressure profile

Figure D1 plots the granular pressure (averaged along the $x$ and $y$ directions) in the system as a function of height ($z$ direction) for different times $\Delta t$. The pressure profile is initially hydrostatic ($\Delta t = 0.2$ s), very rapidly a pressure profile develops that leads to an instability at the interface ($z \sim 100$ mm), i.e. a higher granular pressure in the heavier particle layer ($z > 100$ mm) pushes down onto a lower granular pressure in the lighter particle layer ($z < 100$ mm).

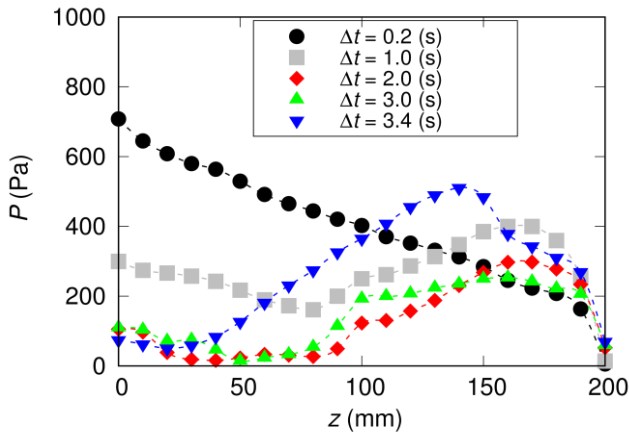

Figure D1: Granular pressure $P$ as a function of the height ($z$) for different instants in time $\Delta t$.